\documentstyle[preprint, aps, epsf, psfig]{revtex}

\begin{document}
\draft
\preprint{}
%\twocolumn[\hsize\textwidth\columnwidth\hsize\csname@twocolumnfalse\endcsname
\title{Consistent Anisotropic Repulsions for Simple Molecules}
\author{Sorin Bastea\cite{sb} and Francis H. Ree\cite{fhr}}
\address{Lawrence Livermore National Laboratory, P. O. Box 808, 
Livermore, CA 94550}
\maketitle
\begin{abstract}
We extract atom-atom potentials from the effective spherical potentials that successfully model Hugoniot 
experiments on molecular fluids, e.g., $O_2$ and $N_2$. In the case of $O_2$ the resulting potentials compare 
very well with the atom-atom potentials used in studies of solid-state properties, while for $N_2$ they are 
considerably softer at short distances. Ground state ($T=0$K) and room temperature calculations 
performed with the new $N-N$ potential resolve the previous discrepancy between experimental and theoretical 
results. 
\end{abstract}
%]
\widetext
In recent years there has been an increasing interest in the high-pressure properties of liquids and 
solids\cite{ashcroft}. As experimental research focused on the `extreme' states of matter, it also forced 
us to refine our knowledge of the relevant intermolecular interactions. In many instances, including the 
study of thermodynamic properties, such interactions can be described by effective two-body potentials. 
Because even simple molecular systems such as $N_2$ or $O_2$ can reveal bewildering complexity at high 
pressures, developing reliable potentials has proven to be a laborious task. A further difficulty is that 
the intermolecular interactions are in most cases anisotropic. Taking into account the anisotropy is 
essential for an accurate description of solid phases. In general, site-interaction models (commonly 
referred to as atom-atom potentials), describing the intermolecular interaction as a sum of the interactions 
between atoms residing on different molecules, have proven to be a successful approach. 

Quantum mechanical calculations have been used to provide information on the intermolecular atom-atom potentials, 
which are further adjusted to reproduce measured quantities such as virial coefficients, sublimation energy, 
melting point at room temperature, low-pressure crystal structures, etc. \cite{etters1,etters2}. These 
experimental data are mainly suitable to constrain the potential in the attractive region. There are also 
abundant high-temperature and high-pressure data which could provide information on the short-range anisotropic 
repulsion. However, they have been mostly interpreted assuming a spherical short-range repulsion, with addition 
of a long-range van der Waals attractive tail \cite{ross1,ross2,ree1}. The main reason is that the statistical 
mechanical theory of particles interacting with spherical potentials is well developed, and reliable and efficient 
computational tools are available. In contrast, such theories for anisotropic interactions are cumbersome, requiring 
approximations that are hard to justify and generally limited in their applicability \cite{hansen}.

The fact that the spherical short-range repulsions employed by earlier workers can successfully explain high-density, 
high-temperature experimental data of molecular fluids suggests that the sphericallization of the anisotropic repulsion 
is reasonable in the high-density, high-temperature environment. In this work we set out to investigate to what extent 
one can extract the anisotropic interactions, in particular at short range, from such spherical interactions. 
Considering the extensive database of spherical intermolecular potentials for both like and unlike pairs, the atom-atom 
potentials so extracted could provide a unified description of pure and mixed molecular solids under a variety of 
pressure and temperature conditions\cite{ashcroft}.

Physically, the spherical potentials extracted from experimental data for anisotropic systems are effective interactions, 
resulting from `averaging' the anisotropy at high temperatures. A great deal of effort has gone into justifying and finding 
procedures to construct such effective spherical interactions. In this regard, the so-called `median' potential of 
Shaw, Johnson, 
and Holian \cite{shaw} represents one outstanding contribution. Subsequently, Lebowitz and Percus \cite{joel1} clarified 
the main ideas behind this approximation, which we now summarize for the slightly more general case of a mixture. 

If $\phi^{ij}(r, \Omega_1, \Omega_2)$ is an anisotropic potential between molecules of type $i$ and $j$, and 
$\phi^{ij}_{0}(r)$ is the corresponding effective spherical potential, a perturbation potential,
\begin{equation} 
\phi^{ij}_{\gamma}(r, \Omega_1, \Omega_2)=
\phi^{ij}_{0}(r)+\Delta_{\gamma}[\phi^{ij}, \phi^{ij}_{0}], 
\end{equation}
is constructed to connect $\phi^{ij}_{0}(r)$ at $\gamma=0 $ to $\phi^{ij}(r, \Omega_1, \Omega_2)$ at $\gamma=1$ 
with $0\leq\gamma\leq 1$, $\Delta_0[\phi^{ij}, \phi^{ij}_0]=0$, $\Delta_1[\phi^{ij}, \phi^{ij}_0]=\phi^{ij} - \phi^{ij}_0$. 
The first-order Helmholtz free energy correction along this path is $\Delta F = \gamma(\partial F/\partial\gamma)|_0$ and 
is proportional to $\sum_{ij}x_i x_j \int g^0_{ij}({\bf r_1}, {\bf r_2}) (\partial\Delta^{ij}_{\gamma}/\partial\gamma)|_0 
d{\bf r_1} d{\bf r_2} d\Omega_1 d\Omega_2$, where $g^0_{ij}$'s are the pair correlation functions of the sphericalized 
system and $x_i$'s are the concentrations ($\sum_{i}x_i = 1$). It is easy to see that $\Delta F$ can be annulled, 
independent of all $g^0_{ij}$'s, by requiring
\begin{equation} 
\int (\partial\Delta^{ij}_{\gamma}/\partial\gamma)|_0 d\Omega_1 d\Omega_2=0
\end{equation}
for all pairs. 

The disadvantage of Eq. (2) as a definition for spherical potentials is that $(\partial\Delta^{ij}_{\gamma}/\partial\gamma)|_0$ 
is, to a large extent, arbitrary. The median is defined by $(\partial\Delta^{ij}_{\gamma}/\partial\gamma)|_0 = 
sgn(\phi^{ij} - \phi^{ij}_0)$ and minimizes the integral of absolute potential deviations, 
$\int|\phi^{ij}-\phi^{ij}_{0}|d\Omega_1 d\Omega_2$ 
\cite{percus}. We also note that, as a consequence of the Gibbs-Bogoliubov inequalities, for the common case 
$\Delta_{\gamma}[\phi^{ij}, \phi^{ij}_{0}]=\gamma[\phi^{ij}-\phi^{ij}_{0}]$ the first order perturbation theory 
provides a rigorous upper bound on the free energy of the system of interest \cite{hansen}. While for the 
general case that we consider here only approximate inequalities hold \cite{sjr}, they do lead to 
useful heuristic variational principles.

One of the main advantages of the median potential over other effective spherical potentials 
\cite{percus} is that it is independent of density and temperature (and also concentrations in the case of mixtures 
\cite{williams,bastea}). Despite its essentially heuristic basis \cite{joel1,macg}, this appears to be an optimal choice 
for the thermodynamics of simple fluids such as $N_2$ and $CO_2$ at high pressures and temperatures \cite{shaw}, where the 
repulsive interactions make dominant contributions. Hence, we treat the spherical potentials derived from experimental 
data (such as Hugoniot data for $N_2$ and $O_2$) as median potentials and develop a procedure to invert them to extract 
the corresponding anisotropic potentials.  

To illustrate the basic idea behind such an inversion process, consider first a simpler system of dumbbells consisting of 
two hard spheres (with diameter $R$) with their centers fixed at length $L$ apart. In this case the equivalent problem 
is to determine $R$ by inverting an effective hard-sphere potential (with diameter $R_m$).
The median definition of $R_m$ is $\int sgn(R_{12} - R_m) d\Omega_1 d\Omega_2=0$ \cite{joel2}, where $R_{12}$ is the 
distance of closest approach between two dumbbells oriented at solid angles $\Omega_1$ and $\Omega_2$. It is fairly simple 
to show that this equation has a unique solution $R$. 

Potentials that are used to model the thermodynamics of high-pressure, high-temperature fluids are more complicated 
than pure hard-cores. The most successful one is the so-called exponential-6 (`exp-6') potential, 
\begin{equation}
v(r) =\frac{\epsilon}{\alpha-6}\left\{6{\tt exp}[\alpha(1-\frac{r}{r_0})]-\alpha\left(\frac{r_0}{r}\right)^6\right\},
\end{equation}
that combines the short-range exponential repulsion found in early quantum-theory calculations \cite{abraham} with 
a truncated dispersive interaction \cite{london} that accounts for the attraction at longer distances. This functional 
form has been found to provide a good representation of the experimental Hugoniots of $N_2$, $O_2$, and other small 
molecules \cite{ross1,ross2,ree1}, while satisfying to a large extent the 'law of corresponding states' \cite{ross1}. 
In this work we use the potential parameters for nitrogen, $\epsilon^{N_2} /{k_B}= 101.9$K, 
$r_0^{N_2} = 4.09$\AA, $\alpha^{N_2} = 13.2$, obtained by Ross \cite{ross1,ross2} and the oxygen parameters, $\epsilon^{O_2} 
/{k_B}= 125$K, $r_0^{O_2} = 3.86$\AA, $\alpha^{O_2} = 13.2$, that have been also successfully used to model the molecular 
dissociation under shock compression \cite{ross2,ree1}. As experimental Hugoniots are excellent probes of the short-range 
repulsive interactions, such potential parameters are perhaps the best representation of the repulsive region under the 
spherical approximation. We also note that the well position $r_0$ and depth $\epsilon$ of the above potentials 
are in very good agreement with the ones obtained in molecular beam scattering experiments \cite{pirani}.

The present inversion process derives an anisotropic potential $\phi(r, \Omega_1, \Omega_2)$ that 
maps into $v(r)$ (with the exp-6 potential parameters described above) through the median construction,
\begin{equation} 
\int sgn[\phi(r, \Omega_1, \Omega_2) - v(r)]d\Omega_1 d\Omega_2=0,
\end{equation}
at all intermolecular distances $r$, where $\phi(r, \Omega_1, \Omega_2)$ is the sum of interactions $\varphi(r_{ij})$
between atoms $i$ and $j$ residing on different molecules,
\begin{equation} 
\phi(r, \Omega_1, \Omega_2) = \sum_{ij}\varphi(r_{ij}),
\end{equation}
We also use an exp-6 form to represent $\varphi(r)$ in Eq. (5) and recast Eq. (4) as a $\chi^2$ minimization. 

To elaborate, for a given set of exp-6  parameters $(\epsilon, r_0, \alpha)$ of $\varphi(r)$  
[thereby a given $\phi(r, \Omega_1, \Omega_2)$] we calculate the spherical potential $\phi^0(r;\epsilon,\alpha, r_0)$ that 
satisfies the median condition, Eq. (4), at a large number of 
intermolecular separations ($\simeq 100$), between $r_{min}$ and a large cutoff separation ($r_{max}$) beyond which 
the potential can be neglected. In practice we choose $r_{min}=2$\AA,  [$v(r_{min})/k_{B}\approx 10^5 $K] and $r_{max}=7$\AA, 
which is sufficient to cover the relevant potential region up to very high densities and temperatures. We then minimize the 
function $\chi^2(\epsilon, r_0, \alpha) = \sum_i[v(r_i)-\phi^0(r_i;\epsilon, r_0, \alpha)]^2/[v(r_i)^2 + \epsilon_*^2]$ to 
determine the exp-6 parameters of $\varphi(r)$, where $\epsilon_*$ is the well depth of $v(r)$ \cite{explain}. 

We carried out such minimizations for $O_2$ and $N_2$, with fixed gas-phase bond lengths of $l_{O_2}=1.20741$\AA\, and 
$l_{N_2}=1.097685$\AA. The resulting potential parameters are:
\begin{itemize}
\item O-O potential: $\epsilon/{k_B}=45.64$K, $\alpha=12.52$, $r_0=3.480$\AA;
\item N-N potential: $\epsilon/{k_B}=34.42$K, $\alpha=12.59$, $r_0=3.773$\AA.
\end{itemize}
 
In Fig. 1 we compare the $O-O$ potential with a potential that is used to study the solid-state properties 
of molecular oxygen \cite{etters1}. The agreement is satisfactory, in particular, at short distances. In Fig. 2 we make a similar 
comparison between the $N-N$ potential  and the so-called Etters potential \cite{etters2}, which is widely used for the study of 
molecular nitrogen. Fig. 2 also shows another potential \cite{bohm}, which was recently used for the study of small $N_{2}$ clusters 
\cite{maillet}. It is clear that the potential that we extracted by inverting the median is much softer at small separations than 
either of these two, the main difference being attributable to the exp-6 parameter $\alpha$ that mediates the stiffness of the repulsion. 

Solid nitrogen has been the subject of numerous experimental 
and theoretical studies in the last decades leading to a good understanding of its phase diagram at lower pressures 
\cite{etters2,mills,olij1,mcmahan,mailhiot}. However, questions remain regarding its high-pressure behavior, i.e., above $\simeq$ 20 to 
30GPa. Some of these questions concern the crystal symmetry and thermodynamics of the high-pressure phases, and others the stability 
of molecular nitrogen with respect to atomic dissociation and its possible transition to a metallic state. But ultimately all 
these questions appear to be related to each other \cite{olij1,moore}.
X-ray diffraction and Raman spectroscopy studies of solid molecular nitrogen have revealed an increasing complexity of the phase 
diagram with pressure, e.g., 
the $\epsilon-N_2$ phase, which is apparently a high-pressure distortion of the $Pm3n$ cubic lattice of $\delta-N_2$ \cite{etters2}. 
Calculations using a site interaction model such as the Etters potential have successfully described the low pressures part of 
the phase diagram. However, they show sharp disagreement with the experimental data at high pressure \cite{olij2}, which in addition 
to a high sensitivity of the high-pressure structures to the intermolecular potential \cite{nose}, emphasizes the need for more 
reliable interactions. To test the $N-N$ potential obtained in this work, we first calculate the $0$K-pressure of $\epsilon-N_2$ 
(rhombohedral $R\bar{3}c$) by minimizing at fixed density the energy with respect to distortions of the $Pm3n$ 
structure consistent with the 
$R\bar{3}c$ symmetry \cite{etters2,lesar}. Because experimental data are available only at room temperature, 
we correct the 0K-pressure, $p(\rho, 0)$, by adding the thermal contribution, $3\gamma\rho k_B T$, at each molecular density 
$\rho$ ($\gamma$ is the Gr\"{u}neisen parameter) \cite{olij2}. In addition, we performed isothermal molecular dynamics 
simulations at room temperature. 

Fig. 3 compares these results with the experimental data of \cite{olij2} and the results obtained with the Etters potential 
\cite{belak}. 
The overall agreement is remarkably good, given the fact that the $N-N$ potential has not been adjusted at all and is a significant  
improvement over the Etters potential calculations. 
This demonstrates the crucial contribution of the short-range repulsion to close-packed high-pressure structures. 
Our $0$K energy minimizations do not show any significant relaxation of the $R\bar{3}c$ structure to a lower symmetry, higher-pressure 
$R3c$ lattice \cite{lesar}. This may be due to the fact that we are only considering structures with eight molecules per unit cell 
and also without inclusion of the quadrupole-quadrupole interaction.  However, such a transformation is believed to occur at a still 
higher pressure and low temperature \cite{schif}.

We further test the agreement of the experimental and theoretical results by looking at the isothermal bulk 
modulus $B=V(\partial P/\partial V)_T$. In order to extract $B$ we fit both the experimental and theoretical results 
with a modified Birch equation, that has been previously used to model the bulk modulus of nitrogen at 
pressures up to about 2GPa\cite{stewart}: 
\begin{equation}
P = P_0 + \frac{3B_0}{2}(y^7-y^5)[1-\xi(y^2-1)]
\end{equation}
where $y=(V_0/V)^\frac{1}{3}$. The comparison shown in Fig. 4 yields very satisfactory agreement.

Local-density-functional energy calculations \cite{mailhiot} yield a reliable description of molecular nitrogen, but one that is 
limited to $0$K. In order to further asses the accuracy of our $N-N$ potential we calculate the $0$K-energy as a function of density 
for $\alpha-N_2$. Fig. 5 shows again good agreement between these data and the corresponding local-density-functional results 
\cite{mailhiot}. The potential obtained in the present work gives a slightly steeper isotherm at high density, producing good agreement 
with experimental pressures in Fig. 3.

We conclude that the $N-N$ and $O-O$ potentials extracted from the median provide a good representation of the short-range 
repulsive region. Given the fact that classical molecular dynamics and Monte Carlo simulations are still 
the most accurate and efficient way of studying anisotropic molecular systems at finite temperatures, the need for reliable potentials 
for such simulations cannot be overstated. The present work has shown that the task of constructing such anisotropic interactions is 
greatly simplified by inverting a spherical potential  based on experimental data. The resulting potential is  `transferable', in that 
it can self-consistently  describe material properties of different solid phases as well as liquid phases. If desired, one can refine 
the attractive part of the potential with inclusion of additional multipolar interactions \cite{etters2}. Such additional refinements
that can effectively capture the effect of the long range two and three-body dispersive interactions may be necessary to 
describe properties at low pressures and temperatures. However, at very high pressures the structure and thermodynamics of the fluid 
is dominated by excluded volume effects well represented by two body potentials \cite{ross1,hoef} that, as we show here, are 
also `transferable'.

We thank Andrew K. McMahan for kindly providing the results published in \cite{mailhiot}. This work was performed under the auspices 
of the U. S. Department of Energy by University of California Lawrence Livermore National Laboratory under Contract No. W-7405-Eng-48.

\begin{figure}
\caption{Repulsive region comparison between the present $O-O$ potential (circles) 
and that of ref. \protect\cite{etters1} (solid line).}
\label{Fig1}
\end{figure}

\begin{figure}
\caption{Repulsive region comparison between the present $N-N$ potential (circles) 
and that of ref. \protect\cite{etters2} (solid line) 
and ref. \protect\cite{bohm} (dashed line).}
\label{Fig2}
\end{figure}

\begin{figure}
\caption{Room temperature isotherm of $N_2$: molecular dynamics (solid line) 
and energy minimization (dot-dashed line) results 
obtained with the $N-N$ potential of this work; Etters potential calculations from 
ref. \protect\cite{belak} (dotted line); experimental data 
of ref. \protect\cite{olij2} (circles: $\epsilon-N_2$, triangles: $\delta-N_2$).}
\label{Fig3}
\end{figure}

\begin{figure}
\caption{Bulk modulus $B$ of $N_2$ extracted from molecular dynamics results 
obtained with the $N-N$ potential of this work (solid line) and experimental data 
of ref. \protect\cite{olij2} (circles: $\epsilon-N_2$, triangles: $\delta-N_2$).}
\label{Fig4}
\end{figure}

\begin{figure}
\caption{$O$K energy calculations for $\alpha-N_2$ with the $N-N$ potential of this work 
(solid line) and the local-density-functional 
results of ref. \protect\cite{mailhiot}.}
\label{Fig5}
\end{figure}

\centerline{\epsfbox{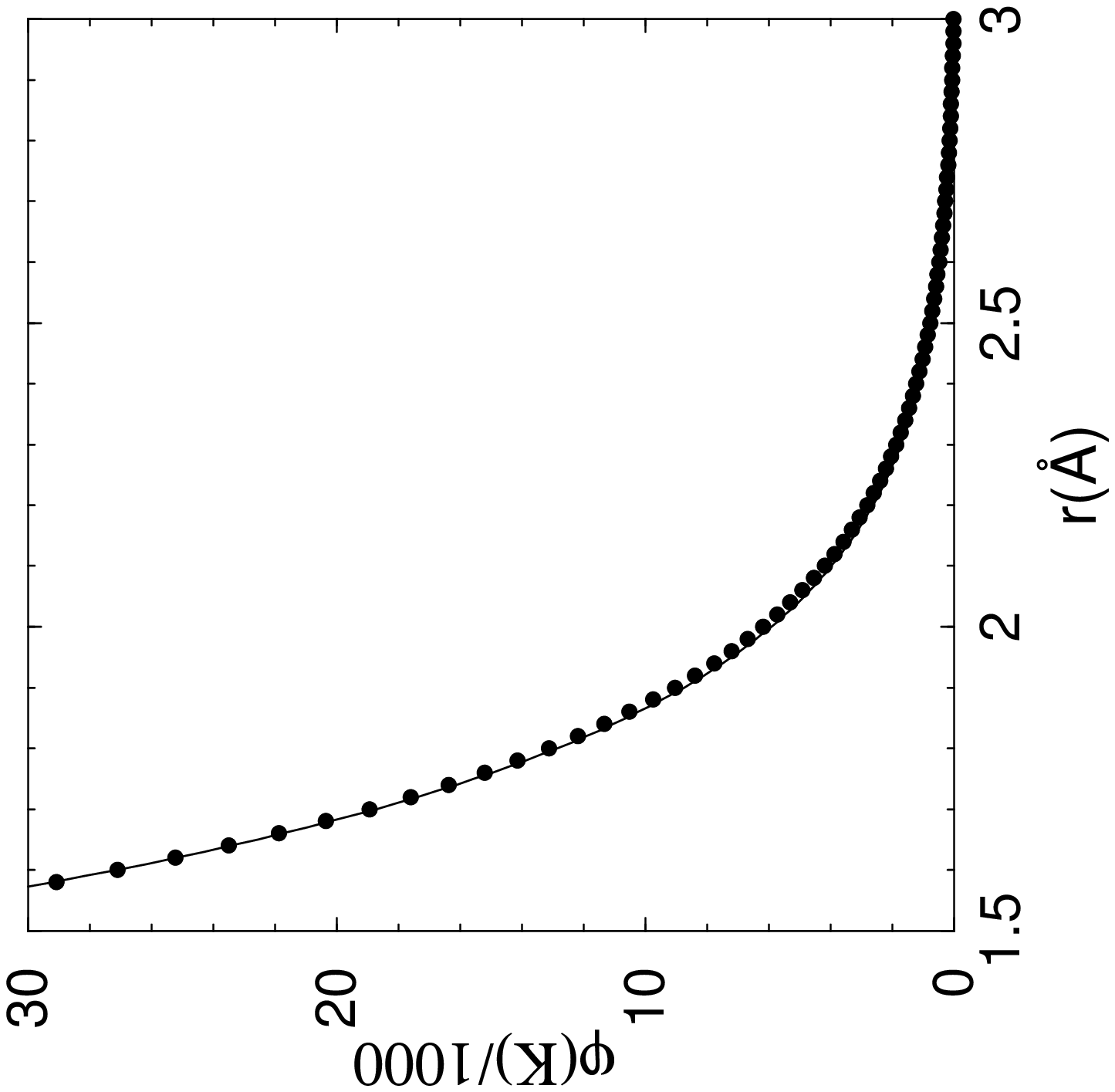}}
\centerline{\epsfbox{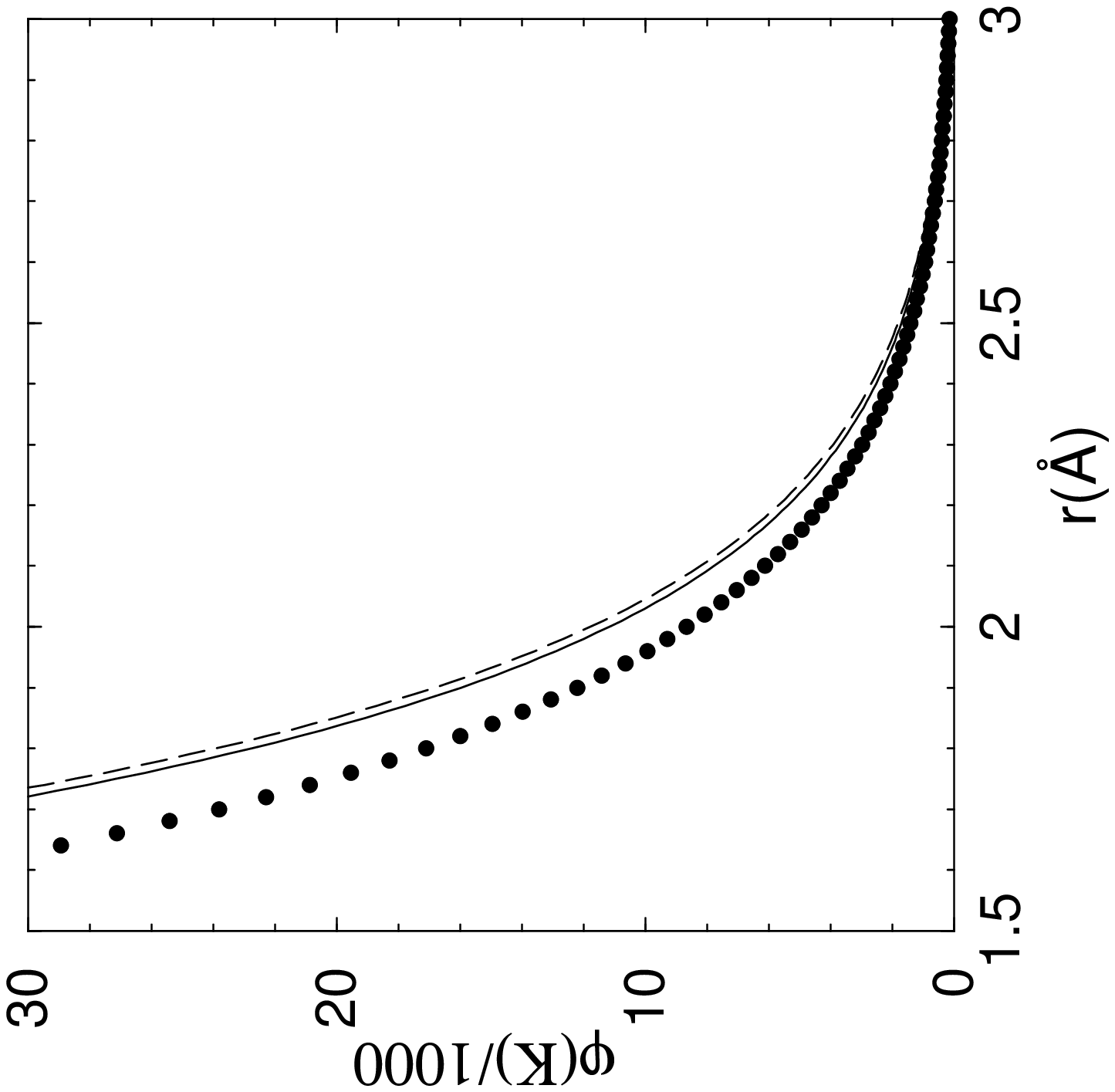}}
\centerline{\epsfbox{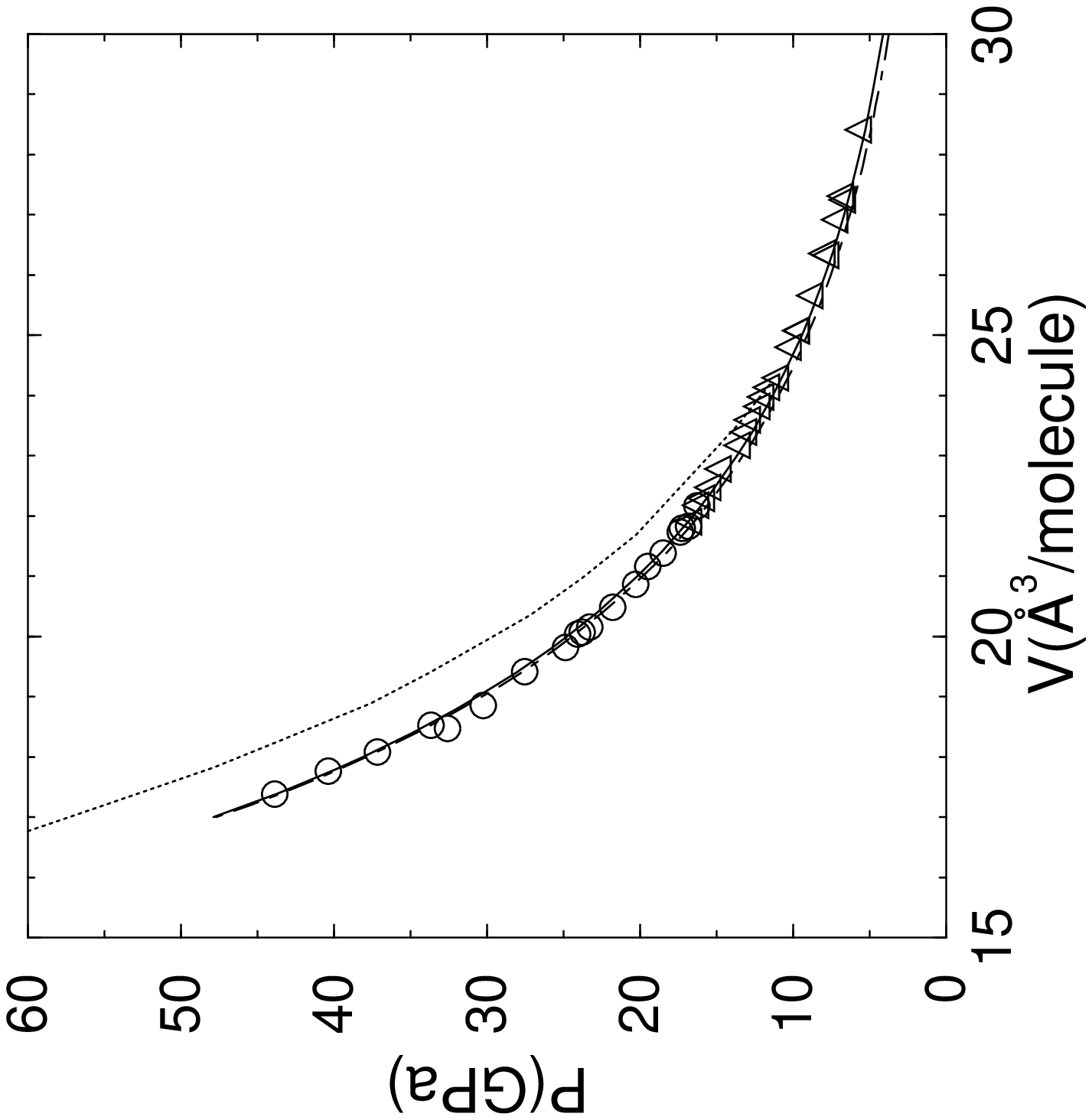}}
\centerline{\epsfbox{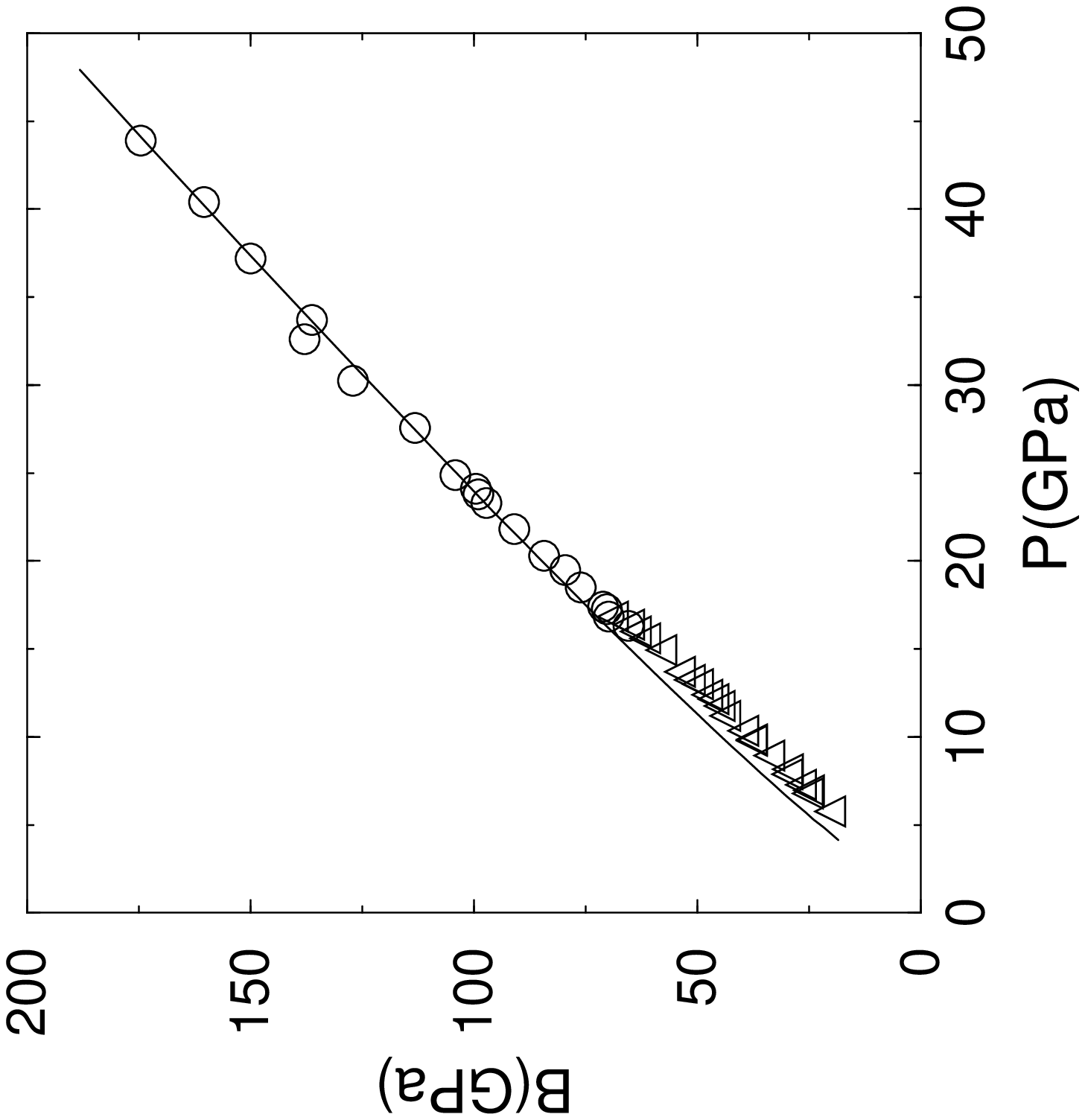}}
\centerline{\epsfbox{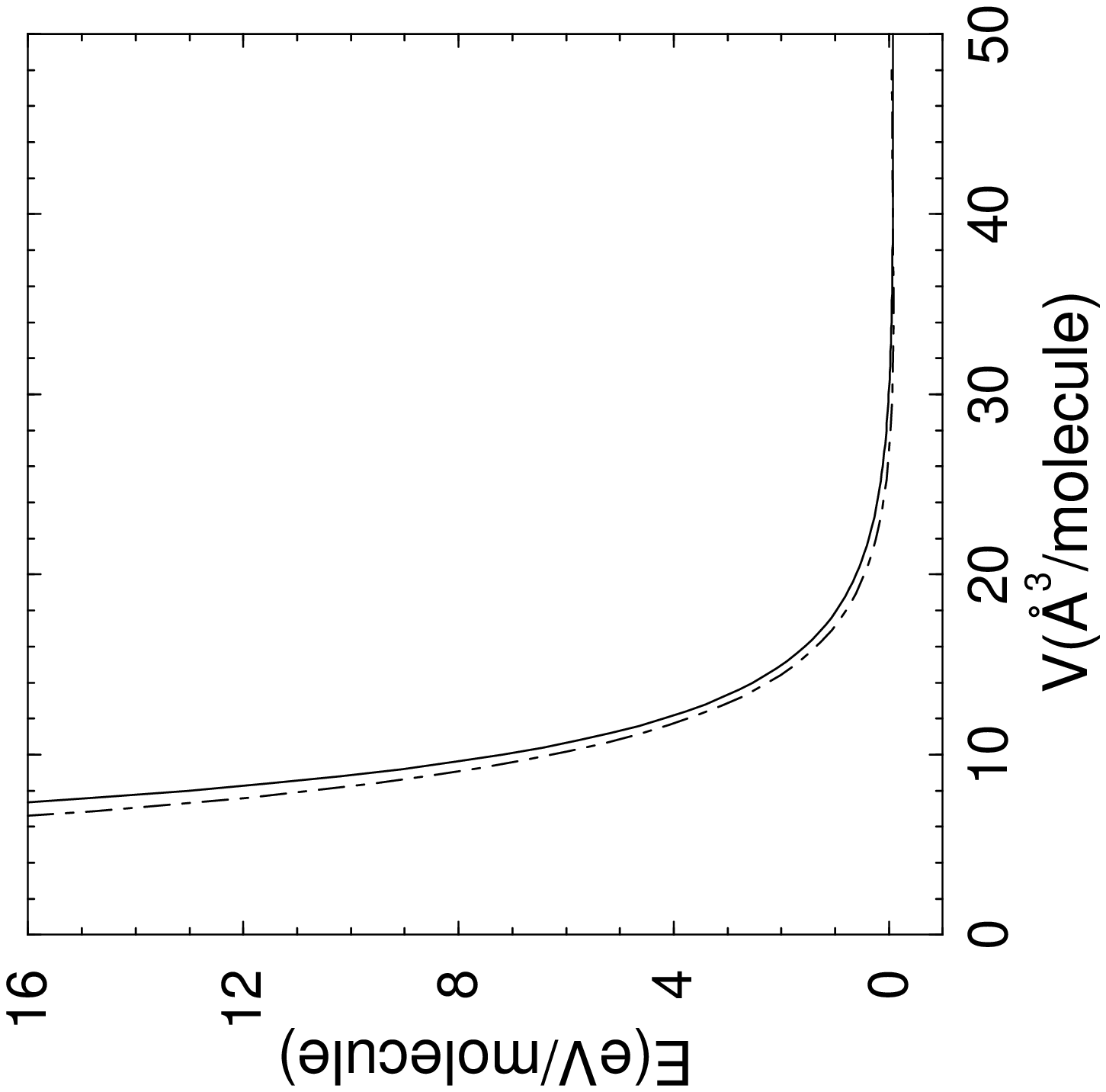}}
%\centerline{\psfig{file=Fig1.ps,width=5.0truein,angle=-90}}
%\centerline{\psfig{file=Fig2.ps,width=5.0truein,angle=-90}}
%\centerline{\psfig{file=Fig3.ps,width=5.0truein,angle=-90}}
%\centerline{\psfig{file=Fig4.ps,width=5.0truein,angle=-90}}
%\centerline{\psfig{file=Fig5.ps,width=5.0truein,angle=-90}}


\begin{references}
\bibitem{explain}We use $[v(r_i)^2 + \epsilon_*^2]$ instead of $v(r_i)^2$ to avoid difficulties at small values of $v(r)$. More complicated forms for $\varphi(r)$, e.g., adding a $1/r^8$-term, while slightly improving the quality of the minimization, lead only to minor changes in the potential. Therefore, for simplicity, we only quote the exp-6 results in this work.
\end{references}
\end{document}